\documentclass[twocolumn,secnumarabic,amssymb, nobibnotes, aps,pra,superscriptaddress]{revtex4-2}
\usepackage{amssymb,amsmath}
\usepackage{graphicx}
\usepackage{subfigure}
\usepackage[english]{babel}
\usepackage{float}
\usepackage{color}
\usepackage[symbol]{footmisc}

\usepackage{graphicx}
\usepackage{natbib}
\usepackage{xcolor}
\usepackage[colorlinks,allcolors=blue]{hyperref}
\usepackage{amsmath}
\usepackage{ulem}
\usepackage{lipsum}

\usepackage{braket}

\begin{document}

\title{Controlling directional propagation in driven-dissipative 2D photonic lattices}

\author{Basti\'an Real}
\email{bastianreal71@gmail.com}
\affiliation{Departamento de F\'isica, Facultad de Ciencias F\'isicas y Matem\'aticas, Universidad de Chile, Santiago, Chile}
\author{Pablo Solano}
\affiliation{Departamento de F\'isica, Facultad de Ciencias F\'isicas y Matem\'aticas, Universidad de Concepci\'on, Concepci\'on, Chile}
\author{Carla Hermann-Avigliano}
\affiliation{Departamento de F\'isica, Facultad de Ciencias F\'isicas y Matem\'aticas, Universidad de Chile, Santiago, Chile}
\affiliation{Millennium Institute for Research in Optics - MIRO, Santiago, Chile}


\begin{abstract}
Controlling light propagation in photonic systems fosters fundamental research and practical application. Particularly, photonic lattices allow engineering band dispersions and tailor transport features through their geometry. However, complete controllability requires external manipulation of the propagating light. Here, we present a resonant excitation scheme to observe quasi-1D and uni-directional propagation of light through the bulk of two-dimensional lattices. To this end, we use the highly anisotropic light propagation exhibited at the energy of saddle points in photonic bands. When multiple drives with judicious amplitudes and phases are tuned to such energy, interference effects between these drives and photonic modes result in controllable directional propagation through the bulk. Similarly, one can formed localized states with controllable localization degrees. We illustrate these effects with driven-dissipative photonic lattices. Our work highlights the importance of external drives for dynamically controlling directional light transport in lattices, a relevant feature for all-optical routing and processing in photonics.

\end{abstract}
\date{\today}

\maketitle


\textit{Introduction.-} Photonic lattices enable the manipulation of light propagation properties. The band structure of a photonic lattice determines the allowed and forbidden frequencies or energies, consequently determining which light waves propagate in certain directions or get confined in a specific volume~\cite{PhCBook}. Photonic lattices are implemented in several experimental platforms with a variety of different periodic geometries, showing their relevance for engineering band structures~\cite{Christodoulides2003,Neshev2007,Szameit2010, Bellec2013, Jacqmin2014}. For example, one- and two-dimensional lattices with a single photonic band have been implemented using coupled waveguides to manipulate light in both the linear and nonlinear regime~\cite{Christodoulides2003, Neshev2007, Szameit2010}. Also, the outstanding dispersive bands of graphene, together with its transport features, have been probed using this same platform~\cite{Song2015}, as well as lattices of coupled microwave resonators~\cite{Bellec2013} and coupled micropillars~\cite{Jacqmin2014, Real2020}. Interestingly, the photonic realm has been a fruitful environment for testing topological phases, whose real-space manifestation is the existence of protected localization and/or unidirectional propagation on the edges~\cite{Wang2009, Hafezi2011, Ozawa2019,Kremer2021,Xia2021}. More recently, theoretical works have predicted exotic quantum dynamics when quantum emitters are coupled to photonic lattices and tuned within the bands, showing directional~\cite{Gonzalez-Tudela2017a,Gonzalez-Tudela2017b} and multi-directional~\cite{Gonzalez-Tudela2019} emission, which has opened up a new route towards harnessing quantum light, even in topologically nontrivial lattices~\cite{Bello2019, Kim2021, Garcia-Elcano2023, Vega2023}.
However, coupling many indistinguishable quantum emitters to 2D photonics lattices is rather challenging, and, to the best of our knowledge, this has yet to be experimentally implemented. A more experimentally accessible way to obtain the same phenomenology is by using a purely optical platform, i.e., dissipative lattices with resonant laser excitations (external drives)~\cite{González-Tudela2022,Jamadi2022}. 

In this work we show how external drives can control light propagation in 2D lattices, from on-demand directional radiation to localization. As a test bed, we use a square lattice resonantly driven by several pumps at the middle-band energy, where the density of states exhibits a van Hove singularity. We find a unitary-driving cell (UDC) that allows observing quasi-1D propagation through the bulk along different directions. An extra control drive modifies the UDC leading to a phase-controlled highly directional propagation through the bulk. Moreover, constructive interference takes place when considering two UDCs and, thus, forming localized states. We finalize showing how to implement these phenomena in other 2D lattices, using a triangular lattice as an example. The experimental realization of our findings could facilitate the external control of all-optical processing of information, such as light routers and logic operations~\cite{Butt2021}.


\begin{figure}[b!]
\centering
\includegraphics[width=\columnwidth]{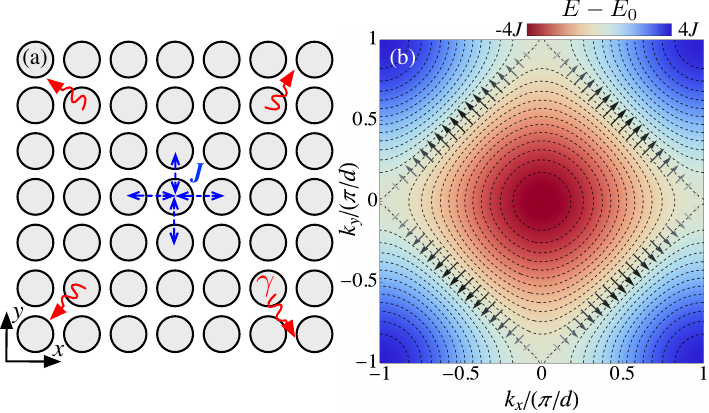}
\caption{(a) Sketch of the square lattice. Gray circles depict the resonators of the lattice, which are coupled to their nearest-neighbors by a strength $J$ (blue double arrows). The losses of light from each resonator is $\gamma=\hbar/\tau$ (red curved arrows). (b) Density map of the dispersion relation for the square lattice in which $E_0$ is the resonant mode energy. Dashed lines show the iso-energies, while vectors display the group velocity at $E=E_0$. $d$ is the nearest-neighbor distance.}
\label{fig1}
\end{figure}

\textit{Tight-binding model and dynamics equation.-} We consider a square lattice of $N$ coupled photonic resonators, as Fig.~\ref{fig1}(a) shows. The light in a resonator can only hop to its nearest-neighbor, thus, the Hamiltonian of this lattice is given by $H_{sq}=-\sum_\mathbf{n,m}J_{\mathbf{m},\mathbf{n}}a^{\dagger}_\mathbf{m}a_\mathbf{n}+\mbox{h.c.}$, where $a^{\dagger}_\mathbf{n}$ ($a_\mathbf{n}$) is the bosonic creation (annihilation) operator in the resonator at the position $\mathbf{n}=(n_xd,n_yd)$ (being $d$ the inter-resonator distance), $J_{\mathbf{m},\mathbf{n}}$ is the nearest-neighbor coupling for $\mathbf{m}\neq \mathbf{n}$, and the on-site energy of resonator mode for $\mathbf{m}=\mathbf{n}$. Considering the discrete Fourier transform $a^{\dagger}_\mathbf{n}=1/\sqrt{N}\sum_\mathbf{k}e^{i\mathbf{k}\cdot\mathbf{r}_\mathbf{n}}a^{\dagger}_\mathbf{k}$ for a homogeneous lattice, i.e., $J_{\mathbf{m},\mathbf{n}}=J$ and $J_{\mathbf{n},\mathbf{n}}=E_0$, the lattice Hamiltonian can be written as $H_{sq}=\sum_\mathbf{k}E(\mathbf{k})a^{\dagger}_\mathbf{k}a_\mathbf{k}$, where $E(\mathbf{k})=E_0-2J\left[\cos(k_xd)+\cos(k_yd)\right]$ is the lattice band. Figure \ref{fig1}(b) presents a density plot of this photonic band covering the energy range $[E_0-4J,E_0+4J]$. Interestingly, at the iso-energy $E=E_0$, there are saddle points where the density of states diverges (van Hove singularity)~\cite{Gonzalez-Tudela2017a,Gonzalez-Tudela2017b,Gonzalez-Tudela2019}. At these points, the group velocity exhibits both zero and its highest values. Specifically, $\mathbf{v}_g(\mathbf{k})=\nabla E(\mathbf{k})=2J(\sin(k_xd),\sin(k_yd))$, which gives that the maximal velocity is $|\mathbf{v}_g|=2J$ at $(k_x,k_y)=\frac{\pi}{d}(0.5,\pm 0.5)$ and $(k_x,k_y)=\frac{\pi}{d}(-0.5,\pm0.5)$, and the minimum one, $|\mathbf{v}_g|=0$, at $(k_x,k_y)=\frac{\pi}{d}(0,\pm 1)$ and $(k_x,k_y)=\frac{\pi}{d}(\pm 1,0)$. Arrows in Fig.~\ref{fig1}(b) represent the amplitude and direction of the group velocity for several $\mathbf{k}$ vectors at $E=E_0$. It is worth highlighting that the extremely anisotropic group velocity in $\mathbf{k}$-space at this iso-energy leads to a highly anisotropic light propagation in real space and, also, a non-Markovian dynamics for quantum emission~\cite{Gonzalez-Tudela2017a,Gonzalez-Tudela2017b}. For any other energy, the group velocity is less anisotropic, whereas it is completely isotropic near $E=E_0 \pm 4J$.

To benchmark this highly anisotropic propagation, we also consider that the resonators can undergo radiative losses and be driven to a specific energy by an external resonant laser. Our theoretical description closely follows experimental implementations with lattices of coupled micropillar~\cite{Rodriguez2016, Jamadi2022}, where temporal dynamics of the light in the square lattice is well modeled by a driven-dissipative set of coupled equations~\cite{Carusotto2013}:

\begin{equation}
i\hbar\frac{\partial \psi_\mathbf{n}}{\partial t}=\left(E_0-i\gamma\right)\psi_\mathbf{n}+\sum_{\mathbf{m}\neq \mathbf{n}} J_{\mathbf{m},\mathbf{n}} \psi_{\mathbf{m}}+F_{\mathbf{n}}e^{-i\omega_{\textrm{d}}t}\,,
\label{eq1}
\end{equation}
where $\psi_{\mathbf{n}}$ represents the amplitude of the resonator field in the $\mathbf{n}$-th position, $\gamma=\hbar/\tau$ is the loss ratio (being $\tau$ the resonator lifetime), and $F_\mathbf{n}$ is the complex amplitude of the resonant excitation laser (drive) at the \textbf{n}-th resonator having photon energy $\hbar\omega_{\textrm{d}}$.


\begin{figure}[t!]
\centering
\includegraphics[width=\columnwidth]{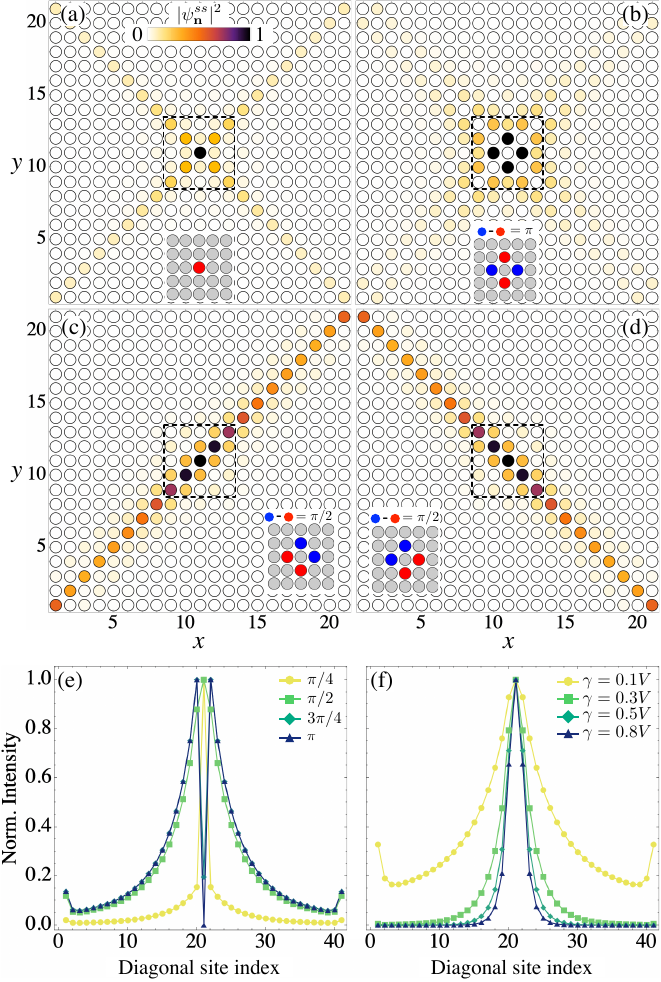}
\caption{\textbf{Quasi-1D propagation}. Steady-state intensity when driving (a) a single lattice site, (b) four sites with alternating $0$ and $\pi$ phases, (c,d) four sites where two have $0$ phase and the other two $\pi/2$, as shown in insets. Dashed squares indicate the zones shown in the insets. $\chi_{\textrm{UDC}}\approx \pm0.8$ for panel (c) and (d), respectively. $N=441$ resonators. (e,f) Profiles of the steady-state intensity along the diagonal sites when resonantly exciting with the UDC in the panel (d) for several values of (e) $\Delta \varphi$ between red and blue spots, and (f) $\gamma$, as indicated on each panel. Intensity profiles in panel (e) and (f) were obtained from simulations with a larger lattice of $N=881$ resonators. Each plot is normalized to its maximal intensity and $\gamma=0.15J$.}
\label{fig2}
\end{figure}
\textit{Numerical results.-}We numerically solve Eq.~(\ref{eq1}) for different drive configurations $F_{\mathbf{n}}$ at an energy $\hbar\omega_{\textrm{d}}$ within the photonic band and calculate the steady-state intensity $|\psi_\mathbf{n}^{ss}|^2$. 

When driving a single resonator at $\hbar\omega_{\textrm{d}}=E_0$, light predominantly propagate along the diagonal resonators, exhibiting an X-type shape, since the resonant drive excites modes with null and maximal velocity. Figure~\ref{fig2}(a) shows this behaviour for $\gamma=0.15J$. The intensity has an exponential decay from the driven resonator to the lattice corners due to losses. For any other energy $E\neq E_0$, light shows some anisotropic propagation along the diagonal because lattice modes have non-zero velocity for every $\textbf{k}$ vector and, conversely, light propagates homogeneously at energies near the maximum and minimum of the band.

Driving multiples lattice resonators yields to interference effects, which have been used to externally engineering localized modes~\cite{Jamadi2022}. In the particular case of the square lattice, at $\hbar\omega_{\textrm{d}}=E_0$, a configuration of four drives with equal amplitudes and phases surrounding a resonator (rhombic shape) gives a highly-localized steady state, which has the entire intensity only on the surrounded lattice site for $\gamma \to 0$~\cite{Jamadi2022,Gonzalez-Tudela2022,Heras2024}.  

We define this rhombic-like spatial arrangement of the drives as the unitary-driving cell (UDC) of the square lattice. Beyond the examples mentioned above, the UDC can also produce extended steady states confined along the diagonals. To demonstrate this, we probe the lattice using such driving configuration with $\hbar\omega_{\textrm{d}}=E_0$. When the drives have equal amplitudes with alternating $0$ and $\pi$ phases, as shown in the inset of Fig.~\ref{fig2}(b) (blue (red) disk depicts $\pi$ ($0$) phase), the steady state displays a destructive interference on the central resonator as well as on the ones along the diagonal (see Fig.~\ref{fig2}(b)). Conversely, the lines of resonators adjacent to the diagonals show a rather intense propagation towards the corners, forming a decaying double X-type of shape. This propagation changes dramatically when the UDC has two consecutive drives with $0$ phase and the other two with $\pi/2$ phase (phase difference of $\Delta \phi=\pi/2$), as shown in Fig.~\ref{fig2}(c) and (d) (see also the insets). We observe a highly-confined propagation along one diagonal evidencing a quasi-$1$D propagation of light through the bulk externally induced by the drives. Therefore, by tuning both the energy and the relative phase of the external drives, either $\textbf{k}=\frac{\pi}{d}(\mp0.5,\pm0.5)$ or $\textbf{k}=\frac{\pi}{d}(\pm 0.5,\pm0.5)$ vectors can be selected so that light can travel only along one of the two diagonal directions.

To quantify the degree of confinement along one diagonal produced by the UDC, we define the parameter $\chi=(I_{\textrm{D}}-I_{\textrm{AD}})/(I_{\textrm{D}}+I_{\textrm{AD}})$, where $I_{\textrm{D}}$ ($I_{\textrm{AD}}$) is the summed intensity along the diagonal resonators from top-right (bottom-right) to bottom-left (top-left). Thus, a fully confined propagation along either diagonal gives as an outcome $\chi^{\textrm{max}}=\pm 0.95$ because both $I_{\textrm{D}}$ and $I_{\textrm{AD}}$ sum the intensity on the central resonator. We compute this parameter for the steady states shown in Fig.~\ref{fig2}(c)-(d), obtaining respectively $\chi_{\textrm{UDC}}\approx \pm0.8$ ($\chi_{\textrm{UDC}}>|0.9|$ when considering the three main diagonal lines), and $\chi_{\textrm{UDC}}\to \chi^{\textrm{max}}$ for $\gamma\to 0$. In contrast, the X-shape propagation produced by the single-resonator drive gives $\chi=0$ for any $\gamma$ since resonators along both diagonals are equally intense.

This quasi-1D propagation occurs under almost any value of $\Delta \phi$ as long as $\hbar\omega_{\textrm{d}}=E_0$, except for $\Delta \phi=0$ at which the localization on one resonator happens~\cite{Jamadi2022}. Figure~\ref{fig2}(e) shows intensity profiles along one of the diagonals as a function of the phase of the UDC. We observe that the main consequence is the constructive or destructive interference at the central resonator, taking place for $\Delta\phi=\pi/2$ (green squares) and $\Delta\phi=\pi$ (blue triangles), respectively. For other values, the profile has a shorter decay  ($\Delta \phi =\pi/4$, yellow circles) or does not present a total destructive interference in the surrounded resonator ($\Delta \phi=3\pi/4$, dark green rhombus). We also notice a slightly higher intensity on sites placed at the ends of the diagonal because the resonant lattice modes are the fastest ones, allowing the light to reach these corner sites and then be reflected back. Considering higher loss rates, light decays faster and this boundary effect can be reduced, as shown in panel~(f). Moreover, the steady states get broader along the orthogonal direction, being less confined and with a lower $\chi_{\textrm{UDC}}$ value ($\sim 0.4$ for $\gamma=J$), yet exhibiting a predominant direction.


\begin{figure}[t!]
\centering
\includegraphics[width=\columnwidth]{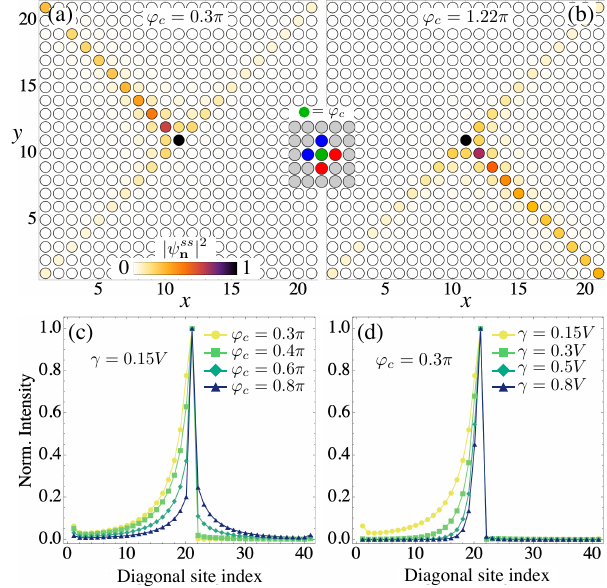}
\caption{\textbf{Harnessing diagonal propagation.} (a)-(b) Steady-state intensity when using the unitary-driving cell (UDC) plus a central drive (see inset) for (a) $\varphi_c=0.3\pi$ and (b) $\varphi_c=1.22\pi$ ($\gamma=0.15J$). $N=441$ sites. (c)-(d) Steady-state intensity profiles along the diagonal resonators for several values of (c) $\varphi_c$ when $\gamma=0.15V$, and (d) $\gamma$ when $\varphi_c=0.3\pi$. $N=881$ sites. The amplitude of the central drive is $|\psi_c|=1.5|\psi_{\textrm{UDC}}^i|$. Each plot is normalized to its maximal intensity.}
\label{fig3}
\end{figure}

The drive-induced quasi-1D propagation can be harnessed so most of the light propagates towards a chosen corner. This is done by adding an extra drive to the lattice site at the center of the UDC. Figure~\ref{fig3}(a) shows the steady-state intensity when the lattice is driven by the UDC plus a central drive with a phase equal to $\varphi_c=0.3\pi$ and an amplitude $1.5$ times the ones of the UDC ($|\psi_c|=1.5|\psi^i_{\textrm{UDC}}|$, $i=1,\dots,4$). The intensity predominantly decays towards the top-left corner of the lattice, whereas null intensity is observed towards the bottom-right corner along the diagonal (see also panel (c), yellow circles). The direction of propagation can be reversed by changing the driving phase of the central site to $\varphi_c=1.22\pi$ (same amplitude), as panel (b) shows. Thus, the control of both the central phase and amplitude enables canceling either wave-vectors $\textbf{k}=\frac{\pi}{d}(0.5,-0.5)$ or $\textbf{k}=\frac{\pi}{d}(-0.5,0.5)$, playing a key role to achieve this quasi-1D directional control. Panel (c) displays the steady-state intensity profiles along the relevant diagonal when driving with the UDC-plus-central configuration. We see that the directional propagation is extinguished for any other value of $\varphi_c$. Importantly, the loss rate increment does not destroy this directional decay along the diagonal, as shown in panel (d) for several values of $\gamma$. Furthermore, some light propagates along the orthogonal diagonal diminishing the quasi-1D confinement. Despite this, for the optimal phase values and $\gamma=0.15J$, $68\%$ of the light is addressed from the center towards the desire corner along the three main diagonals, whereas $13\%$ of the light gets dispersed on the resonators along the opposite one. Defining the parameter $\chi^{\textrm{D}}=(I_l-I_r)/I_{\textrm{AD}}$, where $I_l\,(I_r)$ is the intensity on the resonators along the left-half (right-half) of the diagonal, we quantify a directional degree along the predominant diagonal for these steady states. Thus, $\chi^{\textrm{D}}=0.96$ and $\chi^{\textrm{D}}=-0.96$ for the central-driving phase equal to $\varphi_c=0.3\pi$ and $\varphi_c=1.22\pi$, respectively, being the maximum value equal to the unity. Additionally, $\chi^{\textrm{D}}$ gets much less diminished than $\chi_{\textrm{UDC}}$ when increasing $\gamma$, taking a value of $\chi^{\textrm{D}}=|0.9|$ for $\gamma=J$.

\begin{figure}[ht!]
\centering
\includegraphics[width=\columnwidth]{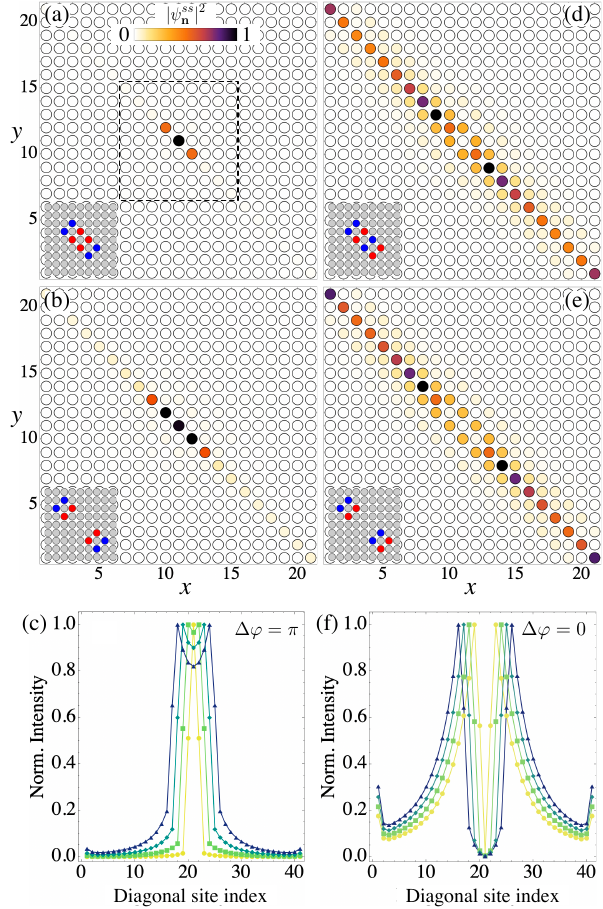}
\caption{\textbf{Engineering localized states with two unitary-driving cells.} Steady-state intensity for (a) $1$ and (b) $3$ resonators in between the drives for $\Delta\varphi=\pi$. Constructive interference occurs in the center of the diagonal. Steady-state intensity for (d) $1$ and (e) $3$ resonators in between the UDCs for $\Delta\varphi=0$. Destructive interference occurs in the center of the diagonal. Simulation in panels (a), (b), (d), and (e) were performed using a lattice with $N=441$ resonators. Each plot is normalized to its maximal intensity. Panels (c) and (f) show the intensity profiles of the steady states along the main lattice diagonal when using two UDCs for several odd separations. Yellow (circles), green (square), dark green (diamond) and blue (triangle) lines correspond respectively to $1$, $3$, $5$, $7$ resonators in between the drives and using a lattice with $N=881$ resonators. Left (right) panels show the case when the two UDCs have a relative phase of $\Delta\varphi=\pi$ ($\Delta\varphi=0$).}
\label{fig4}
\end{figure}

The quasi-1D propagation can also be manipulated to create localized steady states. Simultaneously driving the lattice with two UDCs, separated by a given number of resonators along one diagonal, enables us to control the interference between their independent steady states. Specifically, we consider two identical UDCs as the one shown in the inset of Fig.~\ref{fig2}(d), separated by $i\in \mathbb{N}$ sites along the diagonal. We first consider the drives separated by $i=1$ (see inset in Fig.~\ref{fig4}(a)) and a global phase is added to one of them ($\Delta \varphi$). To obtain a complete scenario of the interference patterns, we sweep $\Delta \varphi$ in the range $[0,2\pi]$. Constructive and destructive interference are observed in the central resonator of the lattice when $\Delta \varphi=\pi$ and $\Delta \varphi=0$, respectively, as shown in Fig.~\ref{fig4}(a) and (d). Remarkably, the constructive interference exhibits a well-localized steady state with intensity on approximately three resonators, and no appreciable intensity on the adjacent lines. We quantify this by computing the inverse participation ratio determined as $\mbox{IPR}=\sum_\mathbf{n}|\psi_\mathbf{n}^{ss}|^4/(\sum_\mathbf{n}|\psi_\mathbf{n}^{ss}|^2)^2$, where $\mbox{IPR}\to1$ ($\mbox{IPR}\to 0$) is a fully localized (delocalized) state. For the single site spacing between the two UDCs, we obtain a steady state value of $\mbox{IPR}_1=0.26$ or, inversely, light occupies only $3.8$ sites. 

The localization along the diagonal can be extended by further separating the drives. Figure~\ref{fig4}(b) displays the intensity profile of a steady state for $i=3$ sites in between the two UDCs. We see that the light is mainly concentrated on the diagonal resonators surrounded by the drives, and decays exponentially towards the corners. Its respective degree of localization is $\mbox{IPR}_{3}=0.12$. This feature can be better seen in Fig.~\ref{fig4}(c), in which the intensity profile of the steady states along the diagonal is plotted for several odd-site separations of the UDCs. In contrast, driving the lattice with the two UDCs in phase ($\Delta\varphi=0$) results in steady states with almost zero light in between the drives, as shown in Fig.~\ref{fig4}(e) for $i=3$ and Fig.~\ref{fig4}(f) for several separations. In the simulations, light tends to accumulate in the corners due to the imposed boundary conditions, which can be reduced by increasing the lattice size. Performing the same phase sweep for even-sites separations of the UDCs, we obtain that localized (delocalized) steady states are formed for opposite value of the phase difference, i.e., $\Delta \phi=0$ ($\Delta \phi=\pi$). In this case, the closest driving configuration is two overlapped UDCs (zero sites in between), which gives a steady state with the highest localization, $\mbox{IPR}_0=0.4$, in which light is mostly located on the two central sites.

\begin{figure*}[t!]
\centering
\includegraphics[width=0.98\textwidth]{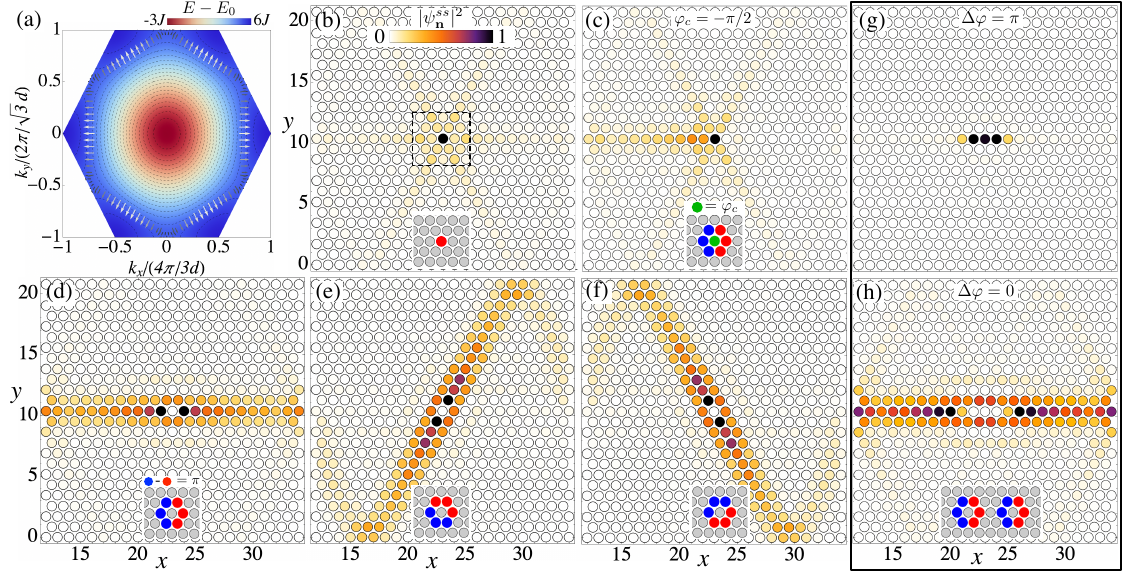}
\caption{\textbf{Directional propagation and localization in a triangular lattice.} (a) Density map of the dispersion relation for the triangular lattice in the first Brillouin Zone. Dashed lines show the iso-energies and vectors depict the group velocity at $E-E_0=-2J$. Vectors depict the group velocity at this energy. Steady-state intensity when the lattice is driven on (b) a single central resonator, and (c) a hexagonal-like drive (UDC) plus an out-of-phase central drive (see inset). (d)-(f) Steady-state intensity when the lattice is driven on six resonators surrounding a single one (hexagonal-like drive), in which three drives have a $0$ phase and the other three have $\pi$ phase (insets display the exact phase distributions). One of the three main directions can be selected depending on the phase distribution. (g) Constructive and (h) destructive interference when two UDCs pump the lattice along one of the main directions. The separation among the drives is one site and their global phase shift is $\Delta\phi=\pi$ and $\Delta\phi=0$, respectively. $N=563$ cavities and $\gamma=0.15J$. Each plot is normalized to its maximal intensity.}
\label{fig5}
\end{figure*}

This directional propagation is not unique of the square lattice, but it is a general feature of 2D lattices with saddle points in their band structures and anisotropic group velocities in the reciprocal space. For example, a triangular lattice has a single photonic band $E(\textbf{k})=E_0-2J[\cos(k_xd)+2\cos(\sqrt{3}k_yd/2)\cos(k_xd/2)]$ with saddle points at $E-E_0=-2J$ ($J$ is the nearest-neighbor coupling). At this energy, there are $\textbf{k}$-dependent maximum and minimum group velocities, as shown in Fig.~\ref{fig5}(a). Therefore, as studied in the square-lattice case, driving the triangular lattice at such an energy by using a single excitation on a central resonator, an anisotropic steady state is obtained with three main directions of propagation (see Fig.~\ref{fig5}(b))~\cite{Navarro2024}.

A unitary-driving cell can be found to control the light propagation along one of these directions and each of them can be deliberately selected by manipulating the phase distribution. Figure~\ref{fig5}(d)-(f) exhibit the intensity of the steady state when using six resonators surrounding a single one,  having three consecutive drives with $0$ phase and the other three with $\pi$ phase (see insets). Depending on the orientation of the phase distribution, specific $\textbf{k}$ vectors can be addressed and, thus, light propagates only along one of the three directions. Moreover, when adding a single drive on the surrounded resonator with $\pm \pi/2$ phase (having, therefore, a hexagonal-like drive plus a central one), the propagation along one half of the main direction is cancelled and enhanced the opposite half, as seen in Fig.~\ref{fig5}(c). 

On the other hand, when simultaneously driving the lattice with two UDCs placed along one of the main directions and separated by one resonator, as shown in Fig.~\ref{fig5}(g)-(h), both constructive and destructive interference occur for an out-of-phase ($\Delta\varphi=\pi$) and an in-phase ($\Delta\varphi=0$) configuration, respectively. In the former case, localized steady states are formed, as seen previously in the square lattice. 

It worth mentioning that all these phenomena is found in more intricate lattice configurations, as long their band structure possesses saddle points. Also, the UDC for each lattice geometry could differ from a given lattice to another one both in amplitude and phase distribution. Nonetheless, at least in all tested lattices, we found that a single resonator must be surrounded by the drives and the phase must be shaped as a dipole-like phase pointed to the direction of propagation (see insets in Fig.~\ref{fig2} and Fig.~\ref{fig5}).   


\textit{Conclusion and Outlook.} We presented a scheme for controlling the directional propagation of light via external drives in two-dimensional lattices of coupled dissipative resonators. We numerically demonstrated that tuning external drives at the energy of saddle points in lattice bands results in an anisotropic propagation, which can be turned into quasi-1D propagation when adding multiple drives with specific amplitude and phase distributions. Similar interference effects allow for manipulating the quasi-1D behavior into highly directional propagation. Lastly, we showed that interference effects between the drives enable the excitation of localized steady states with controllable localization degrees. 

Our numerical results bring to the spotlight the relevance of resonant excitation laser for harnessing the propagation of light in lattices, which can be readily implemented experimentally using, for instance, lattices of coupled micropillars~\cite{Jamadi2022,Pernet2022}. 

We believe that experimental implementations of our findings could find use in all-optical processing of information~\cite{Butt2021}, where simpler versions of phase-controlled optical switches in a 2D photonic crystal have been observed~\cite{Salinas2024}.  

The phenomenology we describe is also relevant in cavity and waveguide QED. From practical application of directional emission in routing photons~\cite{Kannan2023,Solano2023} to fundamental questions in non-Markovian systems~\cite{Gonzalez-Tudela2017a}, where engineering slow group velocities could lead to delay-induce quantum optical effects~\cite{Sinha2020, Alvarez-Giron2024}. 

Moreover, one following avenue of this work is the addition of Kerr-type nonlinearities~\cite{Pernet2022,Heras2024, Usaj2024}, which can expand the degree of control over the localization of light along the predominant directions. In the quantum realm, it would be interesting to study how the quantum properties of squeezed light evolve when the highly directional propagation takes place and how they can be included or implemented in quantum information protocols~\cite{Rojas2019, Medina2021,Rojas2023,ORyan2024}.

\bibliography{sample}{}

\end{document}